\documentclass[11pt]{article}

\usepackage{epsf}
\usepackage{psfig}
\usepackage[dvips]{epsfig}
\usepackage{amsmath}
\usepackage{amssymb}
\usepackage{fullpage}

 \advance \textwidth by -2in
 \advance \textheight by -3in
\def\le{\left(}
\def\ri{\right)}
\def\les{\left[}
\def\ris{\right]}

\def\c#1{\cite{#1}}

\def\epso{\epsilon_o}

\def\muo{\mu_o}
\def\lambdao{\lambda_o}

\def\ko{k_o}
\def\sp{\bf s}
\def\pinc{{\bf p}_+}
\def\pref{{\bf p}_-}

\def\einc{{\bf e}_{inc}(z)}
\def\erefl{{\bf e}_{ref}(z)}
\def\etr{{\bf e}_{tr}(z)}

\def\aL{a_L}
\def\aR{a_R}
\def\rL{r_L}
\def\rR{r_R}
\def\tL{t_L}
\def\tR{t_R}
\def\rRR{r_{RR}}
\def\rRL{r_{RL}}
\def\rLR{r_{LR}}
\def\rLL{r_{LL}}
\def\tRR{t_{RR}}
\def\tRL{t_{RL}}
\def\tLR{t_{LR}}
\def\tLL{t_{LL}}

\def\RLR{R_{LR}}

\def\ux{\hat{\bf{u}}_x}
\def\uy{\hat{\bf{u}}_y}
\def\uz{\hat{\bf{u}}_z}

\begin{document}

\begin{center}
{\large {\bf Theory of electrically controlled exhibition of circular
Bragg phenomenon by an obliquely excited structurally chiral material. Part 2: 
arbitrary dc electric field}}
\vskip 0.5cm

\noindent  { Akhlesh Lakhtakia}\footnote{E--mail: akhlesh@psu,edu}
\vskip 0.2cm
\noindent {\em Computational \& Theoretical Materials Sciences Group (CATMAS)\\
Department of Engineering Science \& Mechanics\\
Pennsylvania State University, University Park, PA 16802--6812, USA}
\vskip 0.5cm

\noindent{Juan Adrian Reyes}\footnote{E--mail: adrian@fisica.unam.mx}
\vskip 0.2cm
{\em Instituto de Fisica\\ 
Universidad Nacional Autonoma de Mexico\\ 
Apartado Postal 20--364, C.P. 01000, Mexico D.F., Mexico}

\end{center}

\vskip 0.5cm

\noindent {\bf Abstract:}
Numerical examination of the solution of
the boundary--value problem of the reflection and transmission of a
plane wave due to a
slab  of an electro--optic structurally chiral material (SCM) indicates
that the exhibition of the circular Bragg phenomenon
by the SCM can be controlled not only by the sign and the magnitude of a
dc electric field but also by its orientation in relation to axis
of helicoidal nonhomogeneity of the SCM. Thereby, the possibility of electrical control of 
circular--polarization filters  has 
been extended. 


\section{Introduction}
In Part 1 \cite{Part1}, we formulated the boundary--value problem of the reflection and transmission of an arbitrary
plane wave due to a
slab  of an electro--optic structurally chiral material (SCM) in terms of
a 4$\times$4 matrix ordinary differential equation. A SCM slab is helicoidally
nonhomogeneous in the thickness direction, and therefore must exhibit the
circular Bragg phenomenon (CBP). Endowed
with one of 20 classes of
point group symmetry, the SCM slab was subjected in Part 1 to a dc electric field  parallel to its axis of nonhomogeneity. The enhancement  of the CBP  by the application of the axial dc electric field has either switching or 
circular--polarization--rejection
applications in optics. The twin possibilities of thinner filters and
electrical control
of the CBP, depending on the local crystallographic class as well as 
the constitutive parameters of the SCM, emerged. 

Our objective here is to generalize the theory of Part 1  to the
application of an arbitrarily oriented dc electric field in order to control the CBP. The matrix ordinary
differential equation then becomes more complicated, even if the
plane wave is normally incident. However, the exhibition of the
CBP is not in doubt, in general, as it depends solely on the structural chirality
of the SCM.

The plan of this paper is as follows: Section \ref{the} contains a brief description
of the optical permittivity matrix of a SCM, and the Oseen transformation is
employed to derive the 4$\times$4 matrix ordinary differential equation. Section 
\ref{numres} contains an account of numerical results and the conclusions
drawn therefrom on the alignment of the dc electric field in relation to the exhibition
of the CBP.

The notation is the same as for Part 1. Vectors are denoted in boldface; the cartesian unit vectors are represented
by $\hat{\mathbf{u}}_x$, $\hat{\mathbf{u}}_y$, and $\hat{\mathbf{u}}_z$;
symbols for column vectors and matrixes are decorated by an overbar; and an $%
\exp(-i\omega t)$ time--dependence is implicit with $\omega$ as the angular
frequency.

\section{Theoretical formulation \label{the}}

We are interested in the reflection and transmission of plane waves due to a
SCM slab of thickness $L$. The axis of helicoidal nonhomogeneity of the SCM is
designated as the $z$ axis, and the SCM is subjected to a uniform dc
electric field $\mathbf{E}^{dc} $. The half--spaces $z\leq 0$ and $z\geq L$
are vacuous. An arbitrarily polarized plane wave is obliquely incident on
the SCM from the half--space $z\leq 0$. As a result, reflected and
transmitted plane waves exist in the half--spaces $z\leq 0$ and $z\geq L$,
respectively. A boundary--value problem has to be solved in order to
determine the reflection and transmission coefficients.

\subsection{Structurally chiral material}

As the electro--optic SCM has the $z$ axis as its axis of helicoidal
nonhomogeneity and is subjected to a dc electric field $\mathbf{E}^{dc}$,
the optical relative permittivity matrix of this material may be stated as 
\begin{equation}
\bar{\epsilon}^{SCM}(z) = \bar{S}_{z}\left(\frac{h\pi z}{\Omega}\right)\cdot%
\bar{R}_{y}(\chi) \cdot\bar{\epsilon}_{PE}(z) \cdot\bar{R}_{y}(\chi)\cdot 
\bar{S}_{z}\left(-\,\frac{h\pi z}{\Omega}\right)\,.  \label{AAepsr}
\end{equation}

The matrix $\bar{\epsilon}_{PE}(z)$ incorporates both the Pockels effect \cite{Boyd}
and the arbitrarily oriented  but uniform $\mathbf{E}^{dc} $. Correct to the first order in the components of the dc electric field, this matrix is given by
\begin{equation}
\displaystyle{\ \bar{\epsilon}_{PE}\approx\left( 
\begin{array}{ccc}
\epsilon _{1}^{(0)}(1-\epsilon _{1}^{(0)}\sum_{K=1}^3 r_{1K}E_{K}^{dc} ) & 
-\epsilon _{1}^{(0)}\epsilon _{2}^{(0)}\sum_{K=1}^3 r_{6K}E_{K}^{dc} & 
-\epsilon _{1}^{(0)}\epsilon _{3}^{(0)}\sum_{K=1}^3 r_{5K}E_{K}^{dc} \\[5pt] 
-\epsilon _{2}^{(0)}\epsilon _{1}^{(0)}\sum_{K=1}^3 r_{6K}E_{K}^{dc} & 
\epsilon _{2}^{(0)}(1-\epsilon _{2}^{(0)}\sum_{K=1}^3 r_{2K}E_{K}^{dc} ) & 
-\epsilon _{2}^{(0)}\epsilon _{3}^{(0)}\sum_{K=1}^3 r_{4K}E_{K}^{dc} \\[5pt] 
-\epsilon _{3}^{(0)}\epsilon _{1}^{(0)}\sum_{K=1}^3 r_{5K}E_{K}^{dc} & 
-\epsilon _{3}^{(0)}\epsilon _{2}^{(0)}\sum_{K=1}^3 r_{4K}E_{K}^{dc} & 
\epsilon _{3}^{(0)}(1-\epsilon _{3}^{(0)}\sum_{K=1}^3 r_{3K}E_{K}^{dc} )
\end{array}
\right) }\,,  \label{PocEps}
\end{equation}
where
\begin{equation}
\left( 
\begin{array}{l}
E_{1}^{dc}(z) \\[5pt] 
E_{2}^{dc}(z) \\[5pt] 
E_{3}^{dc}(z)
\end{array}
\right) = \bar{R}_{y}(\chi)\cdot\bar{S}_{z}\left(-\,\frac{h\pi z}{\Omega}%
\right)\cdot\mathbf{E}^{dc} \,,
\end{equation}
$\epsilon _{1,2,3}^{(0)}$ are the principal relative permittivity scalars in
the optical regime, whereas $r_{JK}$ (with $1\leq J\leq 6$ and $1\leq K\leq
3 $) are the electro--optic coefficients \cite{Part1,Boyd}.
The SCM can be locally  isotropic, uniaxial, or biaxial, depending on the
relative values of $\epsilon_1^{(0)}$, $\epsilon_2^{(0)}$, and $%
\epsilon_3^{(0)}$. Furthermore, the SCM may belong to one of 20
crystallographic classes of local point group symmetry, in accordance with the
relative values of the electro--optic coefficients $r_{JK}$.

The tilt matrix 
\begin{equation}
\bar{R}_{y}(\chi )=\left( 
\begin{array}{ccc}
-\sin \chi & 0 & \cos \chi \\ 
0 & -1 & 0 \\ 
\cos \chi & 0 & \sin \chi
\end{array}
\right)
\end{equation}
involves the angle $\chi \in\left[0,\pi/2\right]$ with respect to the $x$
axis in the $xz$ plane. The use of the rotation matrix 
\begin{equation}
\bar{S}_z(\zeta)=\left( 
\begin{array}{ccc}
\cos \zeta & -\,\sin\zeta & 0 \\ 
\sin\zeta & \cos \zeta & 0 \\ 
0 & 0 & 1
\end{array}
\right)
\end{equation}
in (\ref{AAepsr}) involves the half--pitch $\Omega $ of the SCM along the $z$
axis. In addition, the handedness parameter $h=1$ for structural
right--handedness and $h=-1$ for structural left--handedness. 

Without significant loss of generality, we chose 
\begin{equation}
\mathbf{E}^{dc} = E^{dc} (\hat{\mathbf{u}}_x \cos\chi_{dc} +\hat{\mathbf{u}}%
_z \sin\chi_{dc})\,,\quad \chi_{dc}\in\left[0,\pi/2\right]\,,
\end{equation}
and we note that the case $\chi_{dc}=\pi/2$ has been tackled in Part 1 \cite{Part1}.

\subsection{Propagation in the SCM}

The Maxwell curl postulates for the chosen SCM slab are given by 
\begin{eqnarray}
&&\left. 
\begin{array}{l}
\nabla \times \mathbf{E}(x,y,z)=i\omega\mu_o\mathbf{H}(x,y,z) \\[5pt] 
\nabla \times \mathbf{H}(x,y,z)=-i\omega\epsilon_o\bar{\epsilon}%
^{SCM}(z)\cdot \mathbf{E}(x,y,z)
\end{array}
\right\} \,,  \nonumber \\
&&\qquad\qquad 0<z<L\,,
\end{eqnarray}
where $\epsilon_o$ and $\mu_o$ are the permittivity and the permeability of
free space (i.e., vacuum).

As a plane wave is incident obliquely on the SCM, $\forall z$ we set 
\cite{Part1} 
\begin{equation}
\left. 
\begin{array}{l}
\mathbf{E}(x,y,z)= \mathbf{e}(z)\, \exp\left[
i\kappa(x\cos\phi+y\sin\phi)\right] \\[5pt] 
\mathbf{H}(x,y,z)= \mathbf{h}(z)\, \exp\left[
i\kappa(x\cos\phi+y\sin\phi)\right]
\end{array}
\right\}\,,
\end{equation}
where the wavenumber $\kappa$ and the angle $\phi$ are determined by the
incidence conditions. The essential part of the Maxwell curl postulates can
then be stated in terms of the column vector 
\begin{equation}
{\bar{\psi}}\left( z\right) =\left( 
\begin{array}{c}
e_{x}(z) \\ 
e_{y}(z) \\ 
h_{x}(z) \\ 
h_{y}(z)
\end{array}
\right) \, .  \label{campoe_h}
\end{equation}

As in Part 1\cite{Part1}, it is advantageous to exploit the Oseen transformation  by defining the column vector 
\begin{equation}
{\bar{\psi}}^{\prime }(z)=\bar{M}\left(\frac{h\pi z}{\Omega}\right)\cdot {%
\bar{\psi}}(z)\,,
\end{equation}
where the unitary 4$\times $4 matrix 
\begin{equation}
\bar{M}(\zeta)=\left( 
\begin{array}{cccc}
\cos \zeta & \sin \zeta & 0 & 0 \\ 
-\sin \zeta & \cos \zeta & 0 & 0 \\ 
0 & 0 & \cos \zeta & \sin \zeta \\ 
0 & 0 & -\sin \zeta & \cos \zeta
\end{array}
\right) \,.
\end{equation}
The column vector  ${\bar{\psi}}^{\prime }(z)$ satisfies the 4$\times$4 matrix
ordinary differential equation 
\begin{equation}  \label{oblique}
\frac{d}{dz}{\bar{\psi}}^{\prime }(z)= i \bar{A}^\prime(z)\cdot{\bar{\psi}}%
^{\prime }(z)\,, \qquad 0 < z <L\,,
\end{equation}
where the decomposition 
\begin{equation}
\label{defineA}
\bar{A}^\prime(z) =\bar{A}_0^\prime(u) + \bar{A}_s^\prime(u)\,\sin\chi_{dc}
+ \left[\bar{A}_{cs}^\prime(u)\sin\left(\frac{h\pi z}{\Omega}\right) + \bar{A%
}_{cc}^\prime(u)\cos\left(\frac{h\pi z}{\Omega}\right) \right]\cos\chi_{dc}
\,
\end{equation}
clarifies the significance of the orientation of $\mathbf{E}^{dc}$, and is
correct to the first order in $E^{dc}$.

The various quantities appearing on the right side of (\ref{defineA}) are as follows:
\begin{eqnarray}
\bar{A}_0^\prime(u) &=& \left( 
\begin{array}{cccc}
0 & -i\frac{h\pi}{\Omega} & 0 & \omega\mu_o \\ 
i\frac{h\pi}{\Omega} & 0 & -\omega\mu_o & 0 \\ 
0 & -\omega\epsilon_o\epsilon_2^{(0)} & 0 & -i\frac{h\pi}{\Omega} \\ 
\omega\epsilon_o\epsilon_d & 0 & i\frac{h\pi}{\Omega} & 0
\end{array}
\right)  \nonumber \\[6pt]
&+& \kappa\alpha_3\, \bar{C}_1^\prime(u) +\frac{\kappa^2}{\omega\epsilon_o}\,%
\frac{\epsilon_d}{\epsilon_1^{(0)}\epsilon_3^{(0)}} \,\bar{C}_3^\prime(u)- 
\frac{\kappa^2}{\omega\mu_o} \,\bar{C}_4^\prime(u)\,,
\end{eqnarray}
\begin{eqnarray}
\bar{A}_s^\prime(u) &=&-\,\omega\epsilon_o\frac{\epsilon_2^{(0)}}{%
\epsilon_1^{(0)}} \left( 
\begin{array}{cccc}
0 & \quad0 & \quad0 & \quad0 \\ 
0 & \quad0 & \quad0 & \quad0 \\ 
\epsilon_e+\epsilon_h & \quad -\epsilon_m & \quad0 & \quad 0 \\ 
\epsilon_\iota\cos\chi+(\epsilon_j+\epsilon_\ell) \frac{\sin 2\chi}{2}%
+\epsilon_k\sin\chi & \quad-(\epsilon_e+\epsilon_h) & \quad0 & \quad0
\end{array}
\right)  \nonumber \\[6pt]
&+&\kappa\frac{\epsilon_2^{(0)}}{\epsilon_1^{(0)}\epsilon_3^{(0)}} \left[-\,%
\frac{\alpha_1}{\epsilon_1^{(0)}}\,\bar{C}_1^\prime(u)
+(\epsilon_f+\epsilon_g)\,\bar{C}_2^\prime(u)\right]+ \frac{\kappa^2}{%
\omega\epsilon_o}\, \left(\frac{\epsilon_d}{\epsilon_1^{(0)}\epsilon_3^{(0)}}%
\right)^2\,\frac{\alpha_2}{\epsilon_d}\,\bar{C}_3^\prime(u) \,,
\end{eqnarray}
\begin{eqnarray}
\bar{A}_{cs}^\prime(u)&=& \omega\epsilon_o \left( 
\begin{array}{cccc}
0 & \quad 0 & \quad0 & \quad 0 \\ 
0 & \quad0 & \quad0 & \quad 0 \\ 
-\delta_c & \quad E^{dc}\,\left(\epsilon_2^{(0)}\right)^2 r_{22} & \quad 0 & 
\quad 0 \\ 
\delta_\iota & \quad \delta_c & \quad0 & \quad0
\end{array}
\right)  \nonumber \\[6pt]
&+& \frac{\kappa}{\epsilon_1^{(0)}\epsilon_3^{(0)}} \left[
\delta_j\epsilon_d\,\bar{C}_1^\prime(u)+ \delta_d\epsilon_2^{(0)}\,\bar{C}%
_2^\prime(u)\right] +\frac{\kappa^2}{\omega\epsilon_o}\, \left(\frac{%
\epsilon_d}{\epsilon_1^{(0)}\epsilon_3^{(0)}}\right)^2\,\delta_k\,\bar{C}%
_3^\prime(u)\,,
\end{eqnarray}
\begin{eqnarray}
\bar{A}_{cc}^\prime(u)&=& \omega\epsilon_o \left( 
\begin{array}{cccc}
0 & \quad 0 & \quad0 & \quad 0 \\ 
0 & \quad0 & \quad0 & \quad 0 \\ 
-(\delta_e-\delta_h) & \delta_\ell & \quad 0 & \quad 0 \\ 
\delta_m & \quad \delta_e-\delta_h & \quad0 & \quad0
\end{array}
\right)  \nonumber \\[6pt]
&+& \frac{\kappa}{\epsilon_1^{(0)}\epsilon_3^{(0)}} \left[\delta_n\epsilon_d%
\,\bar{C}_1^\prime(u) + (\delta_f-\delta_g)\epsilon_2^{(0)}\,\bar{C}%
_2^\prime(u)\right] +\frac{\kappa^2}{\omega\epsilon_o}\, \left(\frac{%
\epsilon_d}{\epsilon_1^{(0)}\epsilon_3^{(0)}}\right)^2\,\delta_p\,\bar{C}%
_3^\prime(u)\,, 
\end{eqnarray}
\begin{equation}
\bar{C}_1^\prime(u)= \left( 
\begin{array}{cccc}
\cos u & 0 & 0 & 0 \\ 
-\sin u & 0 & 0 & 0 \\ 
0 & 0 & 0 & 0 \\ 
0 & 0 & \sin u & \cos u
\end{array}
\right)\,,
\end{equation}
\begin{equation}
\bar{C}_2^\prime(u)= \left( 
\begin{array}{cccc}
0 & -\cos u & 0 & 0 \\ 
0 & \sin u & 0 & 0 \\ 
0 & 0 & \sin u & \cos u \\ 
0 & 0 & 0 & 0
\end{array}
\right)\,,
\end{equation}
\begin{equation}
\bar{C}_3^\prime(u)= \left( 
\begin{array}{cccc}
0 & \quad0 & \quad -\sin u\cos u & - \cos^2u \\ 
0 & \quad0 & \quad\sin^2u & \sin u \cos u \\ 
0 & \quad0 & \quad 0 & 0 \\ 
0 & \quad0 & \quad0 & 0
\end{array}
\right)\,,
\end{equation}
\begin{equation}
\bar{C}_4^\prime(u)= \left( 
\begin{array}{cccc}
0 & 0 & \quad 0 & \quad0 \\ 
0 & 0 & \quad0 & \quad 0 \\ 
-\sin u\cos u & - \cos^2u & \quad0 & \quad0 \\ 
\sin^2u & \sin u \cos u & \quad0 & \quad0
\end{array}
\right)\,,
\end{equation}
\begin{eqnarray}
&&\alpha_1= \epsilon _{1}^{(0)}\epsilon_j\cos^2\chi-\epsilon
_{3}^{(0)}\epsilon_\ell\sin^2\chi +\epsilon _{1}^{(0)}\epsilon_k\cos\chi 
\nonumber \\
&&\qquad\quad -\epsilon _{3}^{(0)}\epsilon_\iota\sin\chi\,, \\
&&\alpha_2=\left(\epsilon _{1}^{(0)}\epsilon_n+\epsilon
_{3}^{(0)}\epsilon_p\right)\cos\chi  \nonumber \\
&&\qquad\quad+\left(\epsilon _{1}^{(0)}\epsilon_s+\epsilon
_{3}^{(0)}\epsilon_q\right)\sin\chi\,, \\[6pt]
&&\alpha_3 =\epsilon _{d}\sin 2\chi \frac{\left( \epsilon
_{1}^{(0)}-\epsilon _{3}^{(0)}\right) }{2\epsilon _{1}^{(0)}\epsilon
_{3}^{(0)}}\,,
\end{eqnarray}
\begin{eqnarray}
&&\epsilon _{d}=\frac{\epsilon _{1}^{(0)}\epsilon _{3}^{(0)}}{\epsilon
_{1}^{(0)}\cos ^{2}\chi +\epsilon _{3}^{(0)}\sin ^{2}\chi }\,, \\[9pt]
&&\epsilon_{e} = E^{dc}\, \epsilon_1^{(0)} \epsilon_d
(r_{41}\cos^2\chi-r_{63}\sin^2\chi)\,, \\[9pt]
&&\epsilon_{f}=E^{dc}\,\epsilon_d\sin\chi\,\cos\chi(r_{41}%
\epsilon_3^{(0)}+r_{63}\epsilon_1^{(0)})\,, \\
&&\epsilon_{g} = E^{dc}\, \epsilon_d
(r_{43}\epsilon_3^{(0)}\sin^2\chi+r_{61}\epsilon_1^{(0)}\cos^2\chi)\,, \\
&&\epsilon_{h}=E^{dc}\,
\epsilon_1^{(0)}\epsilon_d\sin\chi\,\cos\chi(r_{43}-r_{61})\,,
\end{eqnarray}
\begin{eqnarray}
&&\epsilon_\iota=E^{dc} \,\frac{\epsilon_1^{(0)}}{\epsilon_2^{(0)}}%
\,\epsilon_d^2 (r_{31}\cos^2\chi-r_{53}\sin^2\chi)\,, \\[9pt]
&&\epsilon_{j}=E^{dc} \frac{\epsilon_1^{(0)}}{\epsilon_2^{(0)}}%
\,\epsilon_d^2 \sin\chi (r_{11} -r_{53} )\,, \\[9pt]
&&\epsilon_{k}=E^{dc}\, \frac{\epsilon_1^{(0)}}{\epsilon_2^{(0)}}%
\,\epsilon_d^2 (r_{13}\sin^2\chi-r_{51}\cos^2\chi)\,, \\[9pt]
&&\epsilon_{\ell}=E^{dc}\, \frac{\epsilon_1^{(0)}}{\epsilon_2^{(0)}}%
\,\epsilon_d^2 \cos\chi (r_{33} -r_{51} )\,,
\end{eqnarray}
\begin{eqnarray}
&&\epsilon_{m}=E^{dc}\,\epsilon_1^{(0)}\epsilon_2^{(0)}
(r_{21}\cos\chi+r_{23}\sin\chi)\,, \\
&&\epsilon_{n} = E^{dc}\, \epsilon_d
(r_{53}\epsilon_3^{(0)}\sin^2\chi+r_{11}\epsilon_1^{(0)}\cos^2\chi)\,, \\
&&\epsilon_{p}=E^{dc}\,\epsilon_d\sin^2\chi\,
(r_{31}\epsilon_3^{(0)}+r_{53}\epsilon_1^{(0)})\,, \\
&&\epsilon_{q} = E^{dc}\, \epsilon_d
(r_{33}\epsilon_3^{(0)}\sin^2\chi+r_{51}\epsilon_1^{(0)}\cos^2\chi)\,, \\
&&\epsilon_{s}=E^{dc}\,\epsilon_d\cos^2\chi\,
(r_{51}\epsilon_3^{(0)}+r_{13}\epsilon_1^{(0)})\,,
\end{eqnarray}
\begin{eqnarray}
\delta_c &=& E^{dc}\, \epsilon_d \,\epsilon_2^{(0)} (r_{42}\cos\chi -
r_{62}\sin\chi)\,, \\
\delta_d&=&E^{dc}\, \epsilon_d
(r_{42}\epsilon_3^{(0)}\,\sin\chi+r_{62}\epsilon_1^{(0)}\,\cos\chi)\,, \\
\delta_e &=& E^{dc} \,\epsilon_d \,\epsilon_2^{(0)}
(r_{43}\cos^2\chi+r_{61}\sin^2\chi)\,, \\
\delta_f &=& E^{dc}\, \epsilon_d
\sin\chi\,\cos\chi\,(r_{43}\epsilon_3^{(0)}-r_{61}\epsilon_1^{(0)})\,, \\
\delta_g&=&E^{dc}\, \epsilon_d
(r_{41}\epsilon_3^{(0)}\,\sin^2\chi-r_{63}\epsilon_1^{(0)}\,\cos^2\chi)\,, \\
\delta_h &=& E^{dc}\, \epsilon_d\,
\epsilon_2^{(0)}\,\sin\chi\,\cos\chi\,(r_{41}+r_{63})\,,
\end{eqnarray}
\begin{eqnarray}
\delta_\iota &=& E^{dc}\,\epsilon_d^2\left[
\sin\chi\,(r_{52}\cos\chi-r_{12}\sin\chi)\right.  \nonumber \\
&&\quad\qquad + \left.\cos\chi\,(r_{52}\sin\chi-r_{32}\cos\chi)\right]\,, \\
\delta_j &=& E^{dc}\,\epsilon_d\left[\epsilon_1^{(0)}
\cos\chi\,(r_{52}\cos\chi-r_{12}\sin\chi) \right.  \nonumber \\
&&\quad\qquad-\left.\epsilon_3^{(0)}
\sin\chi\,(r_{52}\sin\chi-r_{32}\cos\chi)\right]\,, \\
\delta_k&=& E^{dc} \,\left[\epsilon_1^{(0)}
\cos\chi\,(r_{52}\epsilon_3^{(0)}\sin\chi+r_{12}\epsilon_1^{(0)}\cos\chi)
\right.  \nonumber \\
&&\quad\qquad+\left.\epsilon_3^{(0)}
\sin\chi\,(r_{52}\epsilon_1^{(0)}\cos\chi+r_{32}\epsilon_3^{(0)}\sin\chi)%
\right]\,,
\end{eqnarray}
\begin{eqnarray}
\delta_\ell &=& E^{dc}\,\left(\epsilon_2^{(0)}\right)^2\,
(r_{23}\cos\chi-r_{21}\sin\chi)\,, \\
\delta_m &=& E^{dc}\,\epsilon_d^2\, \left[\sin^2\chi\,
(r_{11}\sin\chi-r_{13}\cos\chi)\right.  \nonumber \\
&&\quad\qquad+ \cos^2\chi\,(r_{31}\sin\chi-r_{33}\cos\chi)  \nonumber \\
&&\quad\qquad\left.-2\sin\chi\cos\chi\,(r_{51}\sin\chi-r_{53}\cos\chi)\right]
\\
\delta_n &=& E^{dc}\,\epsilon_d\, \left[\sin^2\chi\cos\chi\,
(r_{11}\epsilon_1^{(0)}-r_{31}\epsilon_3^{(0)})\right.  \nonumber \\
&&\quad\qquad-\sin\chi\cos^2\chi\,
(r_{13}\epsilon_1^{(0)}-r_{33}\epsilon_3^{(0)})  \nonumber \\
&&\quad\qquad\left.-(r_{51}\sin\chi-r_{53}\cos\chi)(\epsilon_1^{(0)}\,\cos^2%
\chi-\epsilon_3^{(0)}\,\sin^2\chi)\right]\,, \\
\delta_p&=&-E^{dc}\left[
\left(\epsilon_1^{(0)}\cos\chi\right)^2(r_{11}\sin\chi-r_{13}\cos\chi)
\right.  \nonumber \\
&&\quad\qquad+
\left(\epsilon_3^{(0)}\sin\chi\right)^2(r_{31}\sin\chi-r_{33}\cos\chi) 
\nonumber \\
&&\quad\qquad\left.+2\epsilon_1^{(0)}\epsilon_3^{(0)}\sin\chi\cos\chi\,
(r_{51}\sin\chi-r_{53}\cos\chi)\right]\,,
\\
u&=& \frac{h\pi z}{\Omega}\,-\phi\,.
\end{eqnarray}

By virtue of linearity, the solution of the 4$\times$4 matrix ordinary
differential equation (\ref{oblique}) must be of the form 
\begin{equation}  \label{oblique-soln1}
{\bar{\psi}}^{\prime }(z_2)= \bar{U}^\prime(z_2-z_1)\cdot{\bar{\psi}}%
^{\prime }(z_1)\,,
\end{equation}
whence 
\begin{eqnarray}
\bar{\psi}(z_2)&=&\bar{M}\left(-\,\frac{h\pi z_2}{\Omega}\right)\cdot\bar{U}%
^\prime(z_2-z_1)\cdot\bar{M}\left(\frac{h\pi z_1}{\Omega}\right)\cdot\bar{%
\psi}(z_1)  \nonumber \\[4pt]
&\equiv&\bar{U}(z_2-z_1)\cdot\bar{\psi}(z_1)\,,  \nonumber \\
&&\qquad \quad0\leq z_\ell\leq L\,,\quad \ell=1,2\,.  \label{oblique-soln2}
\end{eqnarray}
Just as for Part 1 \cite{Part1},
we chose to implement the piecewise homogeneity
approximation method \cite{LakhtakiaB}  to calculate $\bar{U}^\prime(z)$.

\subsection{Reflection and transmission}

The incident plane wave is  delineated by the electric field phasor  
\begin{equation}
\einc= 
\le \aL\,\frac{i\sp-\pinc}{\sqrt{2}} -
\aR\,\frac{i\sp+\pinc}{\sqrt{2}} \ri\,e^{i\ko z\cos\theta}
\,,
\qquad  z \leq 0\,,
\label{eq9.50}
\end{equation}
where 
 $\aL$ and $\aR$ are the
amplitudes of the  LCP
and RCP components, respectively. 
The electric field phasors associated with
the reflected and 
transmitted plane waves, respectively,  
are given as  
\begin{equation}
\erefl= 
\le -\rL\,\frac{i\sp-\pref}{\sqrt{2}} +
\rR\,\frac{i\sp+\pref}{\sqrt{2}} \ri\,e^{-i\ko z\cos\theta}
\,\qquad z \leq 0\,,
\label{eq9.53}
\end{equation}
and
\begin{equation}
\etr= 
\le \tL\,\frac{i\sp-\pinc}{\sqrt{2}} -
\tR\,\frac{i\sp+\pinc}{\sqrt{2}} \ri\,e^{i\ko (z-L)\cos\theta}
\,,\qquad
\quad z \geq L\,.
\label{eq9.54}
\end{equation}
The amplitudes  $r_{L,R}$ and $t_{L,R}$ indicate the as--yet unknown strengths
of the LCP and RCP  components of
the reflected and transmitted plane waves, both of which are
elliptically polarized in general.

The propagation vector of the
 incident  
plane wave 
makes
an angle $\theta \in \les 0,\,\pi/2\ri$ with respect to the $+z$ axis,
and is inclined to the $x$ axis in the $xy$ plane
by an angle $\psi \in\les 0,\,2\pi\ris$; accordingly, the transverse
wavenumber 
$
\kappa = \ko\,\sin\theta$,
where $\ko=\omega\sqrt{\epso\muo}$ is the wavenumber in free space.
The free--space wavelength is denoted by $\lambdao=2\pi/\ko$.
The vectors
\begin{eqnarray}
\label{eq9.51}
&&\sp=-\ux\sin\phi + \uy \cos\phi\,,
\\
\label{eq9.52}
&&{\bf p}_\pm=\mp\le \ux \cos\phi + \uy \sin\phi \ri \cos\theta + 
\uz \sin\theta\,
\end{eqnarray}
are of unit magnitude. 

The reflection--transmission problem amounts to
four simultaneous, linear algebraic equation \cite{Part1,LakhtakiaB}, which  can be solved by standard matrix 
manipulations.
It is usually convenient
to define reflection and transmission coefficients, which 
appear as the elements of the 2$\times$2 matrixes in the following
relations:
\begin{equation}
\label{eq9.55}
\les\begin{array}{c}\rL\\\rR\end{array}\ris
=
\les\begin{array}{cc}\rLL & \rLR\\\rRL & \rRR\end{array}\ris
\,
\les\begin{array}{c}\aL\\\aR\end{array}\ris\,,
\end{equation}
\begin{equation}
\label{eq9.56}
\les\begin{array}{c}\tL\\\tR\end{array}\ris
=
\les\begin{array}{cc}\tLL & \tLR\\\tRL & \tRR\end{array}\ris
\,
\les\begin{array}{c}\aL\\\aR\end{array}\ris\,.
\end{equation}
Co--polarized coefficients have both subscripts identical, but
cross--polarized coefficients do not. The square of the magnitude
of a reflection or transmission coefficient is the corresponding
reflectance or transmittance;  thus, $\RLR = \vert\rLR\vert^2$ is
the reflectance corresponding to the reflection coefficient $\rLR$,
and so on.

\section{Numerical results and conclusion}\label{numres}
With respect to the orientation of ${\bf E}^{dc}$, the right
side of (\ref{defineA}) can be divided into three parts. The first
part is indifferent to ${\bf E}^{dc}$ and therefore to $\chi_{dc}$,
the second shows itself at maximum advantage for \emph{axial}
dc electric fields (i.e., when $\chi_{dc}=90^\circ$),
whereas the third is most effective for \emph{transverse} dc electric fields
(i.e., when $\chi_{dc}=0^\circ$). The effects of the first part have been studied
extensively already \cite{LakhtakiaB}, and those of the second part have
been the focus of Part 1  as well as of other papers \c{RL06,RLno2}.

When considering the effects of the third part as well as the interplay of the second and the
third parts, we must keep in mind that the number of variables for a comprehensive
parametric study is large. These variables include the local isotropy, uniaxiality, or
biaxiality, as determined by the relative values of $\epsilon_{1,2,3}^{(0)}$;
the local point group symmetry of which there are 20 classes, as determined
by the relative values of $r_{JK}$; the two angles of incidence $\theta$
and $\phi$; the angle $\chi$ of the tilt dyadic, the half--pitch $\Omega$, and the normalized
thickness $L/\Omega$; and the angle
$\chi_{dc}$. Given this plethora of variables, we had to restrict the scope
of our investigation.

With guidance from the results reported for Part 1,
we chose to focus on a locally biaxial SCM, since such materials
can offer high electro--optic coefficients which would lower the magnitude
of the applied dc electric field. In particular, we opted for 
the orthorhombic $mm2$ class, choosing the relative permittivity scalars and the 
electro--optic coefficients the same as for potassium niobate \cite{ZSB}. Furthermore, normal incidence  is
 the most common condition for using planar optical devices, and so we
 set $\theta=0^\circ$. Finally, the effect of
 $\phi$ not being significant on the exhibition of the CBP \cite{Part1}, we set
 $\phi=0^\circ$.

\begin{figure}[!ht]
\centering \psfull
\epsfig{file=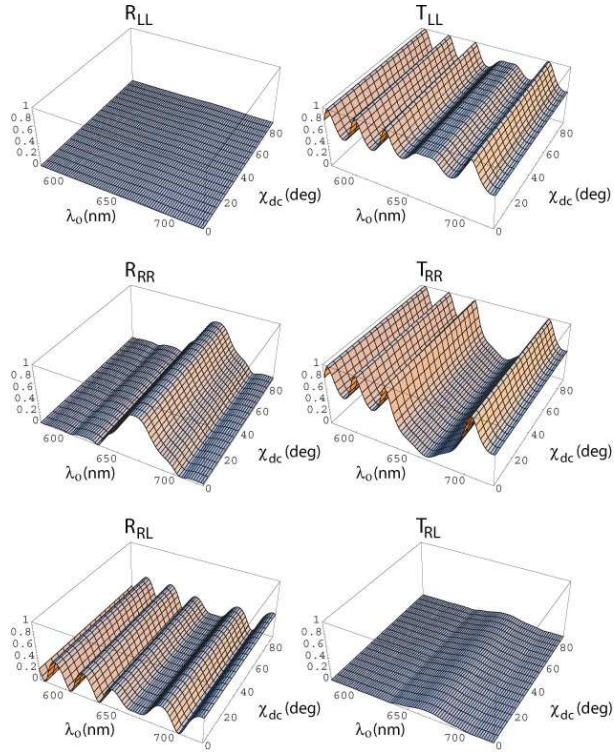,width=8cm }
\caption{Reflectances and transmittances of a locally biaxial SCM slab of thickness
$L=20\,\Omega$ as functions of the free--space wavelength $\lambda_o$
and the orientation angle $\chi_{dc}$ of the
applied dc electric field, when $E^{dc}=10^7$~V~m$^{-1}$ and $\theta=\phi=0^\circ$. The local crystallographic
class 
of the SCM is orthorhombic $mm2$. Other parameters are: $\epsilon_1^{(0)} = 4.72$,
$\epsilon_2^{(0)}= 5.20$,
$\epsilon_3^{(0)}=5.43$, 
$r_{13}=34\times 10^{-12}$~m~V$^{-1}$,
$r_{23}=6\times 10^{-12}$~m~V$^{-1}$,
$r_{33}=63.4\times 10^{-12}$~m~V$^{-1}$,
$r_{42}=450\times 10^{-12}$~m~V$^{-1}$,
$r_{51}=120\times 10^{-12}$~m~V$^{-1}$,
all other $r_{JK}=0$,
$h=1$, $\Omega=150$~nm, and $\chi=90^\circ$. As   $T_{LR}=T_{RL}$   and $R_{LR}=R_{RL}$ to numerical accuracy, the plots of  $T_{LR}$ and
 $T_{LR}$ are not shown.}
\label{orthomm2-1}
\end{figure}

Figure \ref{orthomm2-1} shows the reflectances and transmittance spectrums
of a structurally right--handed SCM with half--pitch $\Omega=150$~nm and
tilt angle $\chi=90^\circ$, when
$E^{dc}=10^7$~V~m$^{-1}$ and $\chi_{dc}\in \left[0^\circ,90^\circ\right]$. No dependence
on $\chi_{dc}$ in the six plots presented 
actually indicates that the magnitude of the dc electric field
is too low to have any significant effect; indeed, the spectrums are virtually the same
as for $E^{dc}=0$. The high ridge in the plot of $R_{RR}$ located at $\lambdao\approx
667$~nm, and its absence
in the plot of $R_{LL}$, are signatures of the CBP, along with the trough
in the plot of $T_{RR}$.

\begin{figure}[!ht]
\centering \psfull
\epsfig{file=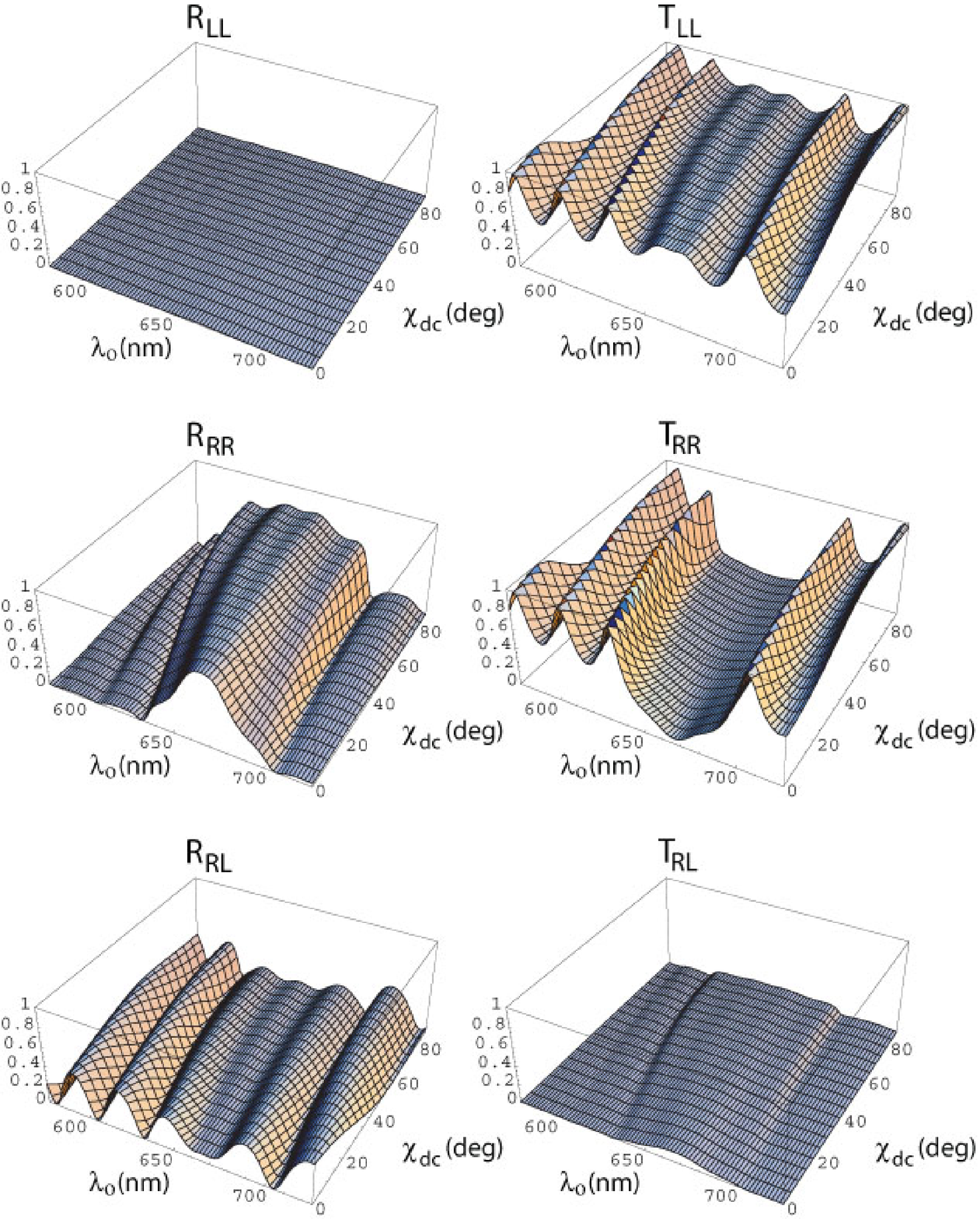,width=8cm }
\caption{Same as Fig.~\ref{orthomm2-1}, except that $E^{dc}=0.67\times10^9$~V~m$^{-1}$. }
\label{orthomm2-2}
\end{figure}

Figure~\ref{orthomm2-2} contains the same plots as the previous figure,
but for $E^{dc}=0.67\times10^9$~V~m$^{-1}$~---~the same value
as used for Fig.~8 of Part 1. This magnitude is high enough to have an effect
on the CBP, which also means that the reflectance and the transmittance
spectrums change with $\chi_{dc}$. The center--wavelength of the
Bragg regime is 646~nm and the full--width--at--half--maximum
bandwidth is 69~nm for $\chi_{dc}=90^\circ$, but the corresponding
quantities are 667~nm and 40~nm for $\chi_{dc}=0^\circ$. In addition,
the peak value of $R_{RR}$ diminishes by about 10\% as $\chi_{dc}$
changes from $90^\circ$ to $0^\circ$.

\begin{figure}[!ht]
\centering \psfull
\epsfig{file=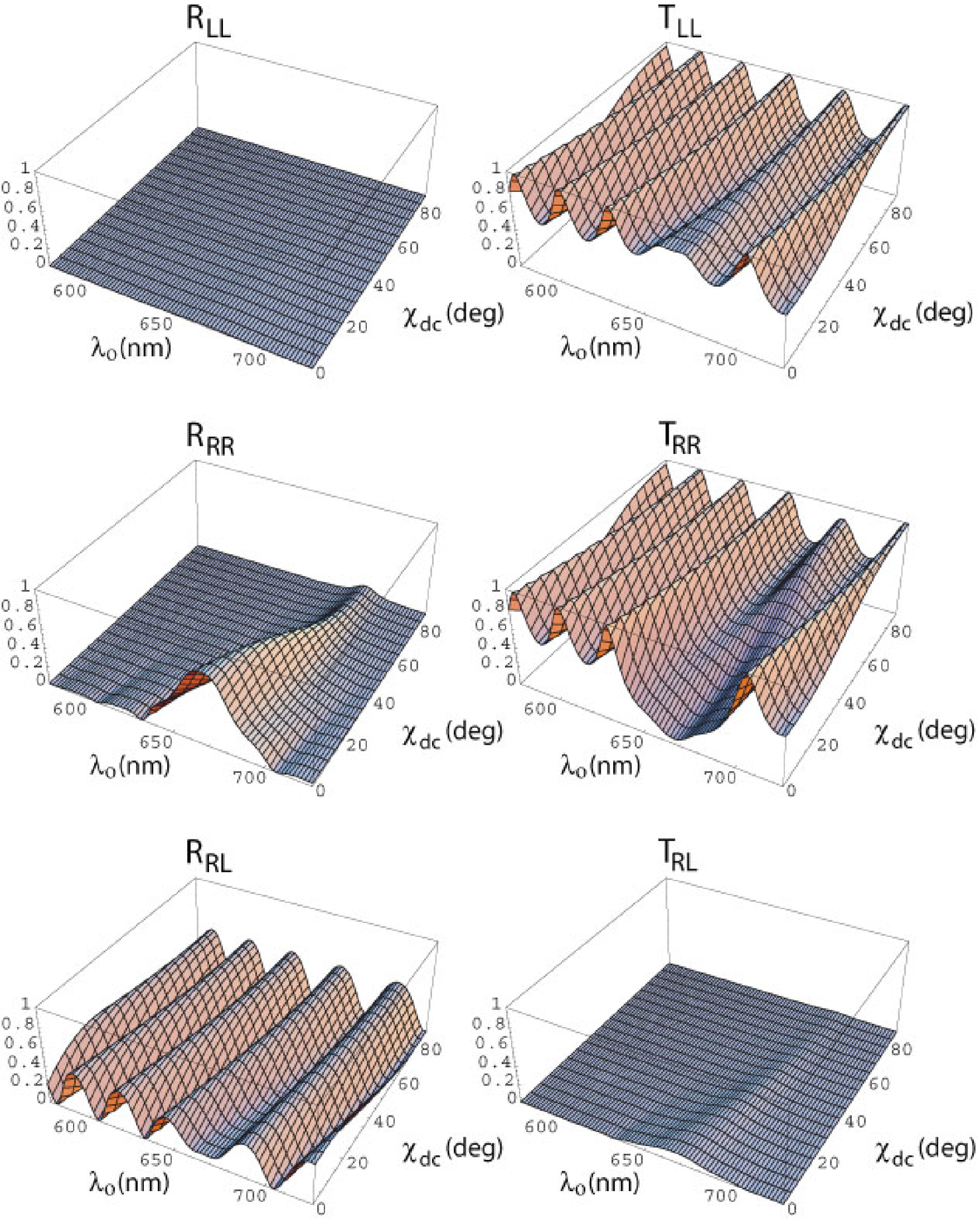,width=8cm }
\caption{Same as Fig.~\ref{orthomm2-1}, except that $E^{dc}=-0.67\times10^9$~V~m$^{-1}$. }

\label{orthomm2-3}
\end{figure}

The situation changes significantly when the sign of $E^{dc}$ is altered,
as exemplified by Fig.~\ref{orthomm2-3} for $E^{dc}=-0.67\times10^9$~V~m$^{-1}$.
The center--wavelength of the
Bragg regime is 688~nm and the full--width--at--half--maximum
bandwidth is 15~nm for $\chi_{dc}=90^\circ$, but the corresponding
quantities remain at 667~nm and 40~nm for $\chi_{dc}=0^\circ$. In addition,
the peak value of $R_{RR}$ increases by about 600\% as $\chi_{dc}$
changes from $90^\circ$ to $0^\circ$. Thus, the exhibition of the CBP
is affected dramatically in the center--wavelength, the bandwidth, and
the peak co--handed and co--polarized reflectance by the sign of $E^{dc}$
as well as the orientation angle $\chi_{dc}$.

\begin{figure}[!ht]
\centering \psfull
\epsfig{file=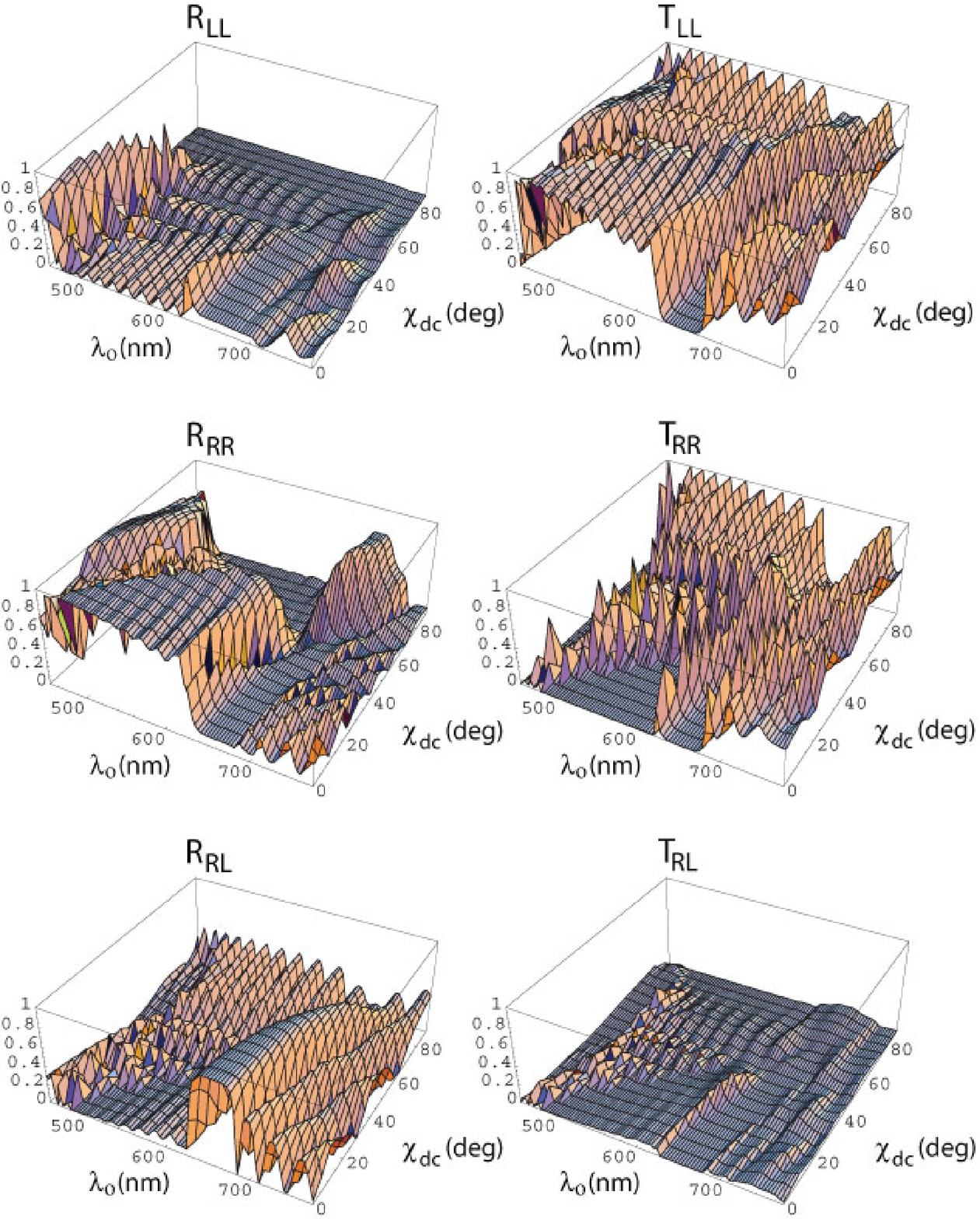,width=8cm }
\caption{Same as Fig.~\ref{orthomm2-1}, except that $\chi=45^\circ$
and $E^{dc}=0.67\times10^9$~V~m$^{-1}$. }
\label{orthomm2-4}
\end{figure}

\begin{figure}[!ht]
\centering \psfull
\epsfig{file=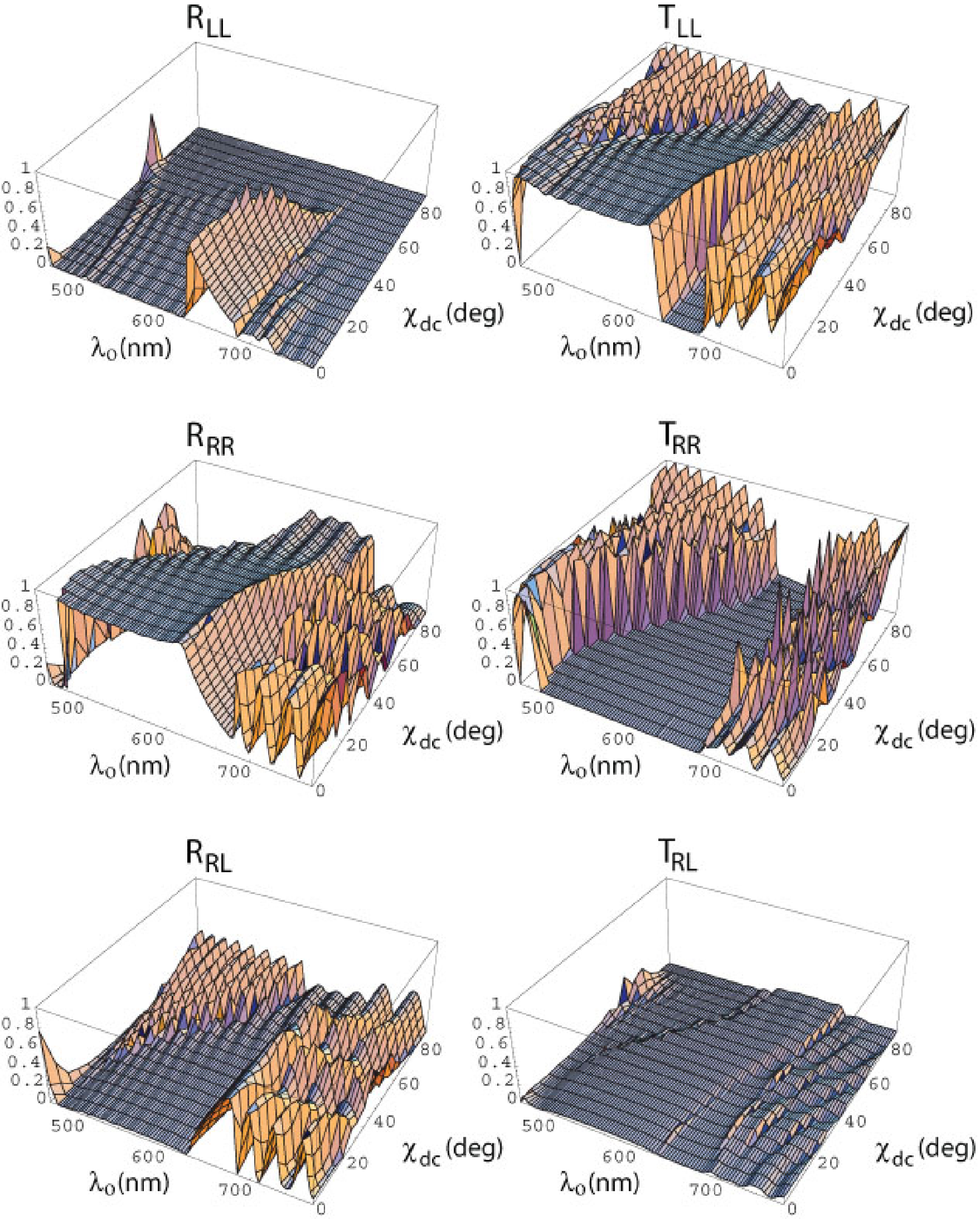,width=8cm }
\caption{Same as Fig.~\ref{orthomm2-1}, except that $\chi=45^\circ$
and $E^{dc}=-0.67\times10^9$~V~m$^{-1}$. }
\label{orthomm2-5}
\end{figure}

Whereas Figs.~\ref{orthomm2-2} and \ref{orthomm2-3} were drawn 
for SCMs with $\chi=90^\circ$,
calculations for Figs.~\ref{orthomm2-4} and \ref{orthomm2-5}
were made for $\chi=45^\circ$. These two figures indicate
a blue--shifting of the CBP on the order of 100~nm as $\chi_{dc}$
changes from $90^\circ$ to $0^\circ$.
Furthermore, the bandwidth is greatly affected by
the value of $\chi_{dc}$ and the sign of $E^{dc}$; indeed,
the CBP vanishes for $\chi_{dc}$ in the neighborhood
of $50^\circ$ when $E^{dc}=0.67\times10^9$~V~m$^{-1}$. Thus, the exhibition
of the CBP is in two different ranges of $\chi_{dc}$ that do not overlap but are
in proximity of each other.

Other types of Bragg phenomenons may appear in the spectral response
characteristics. For example,
Fig.~\ref{orthomm2-4} shows a high--$R_{RL}$ ridge
which suggests that the electro--optic SCM can be made to
function like a normal mirror (high $R_{RL}$ and $R_{LR}$)
in a certain spectral regime than like a structurally right--handed
chiral mirror (high $R_{RR}$ and
low $R_{LL}$) \cite{LX}.

We conclude that the exhibition of the circular Bragg
phenomenon by an electro--optic structurally chiral material
can be controlled not only by the sign and the magnitude of a
dc electric field but also by its orientation in relation to axis
of helicoidal nonhomogeneity. Although we decided to present numerical
results here only for normal incidence, several numerical
studies confirm that our conclusions also apply to 
oblique incidence. Thus, the possibility of electrical control of 
circular--polarization filters, that emerged in Part 1, has 
been reaffirmed and extended. Theoretical studies on particulate
composite materials with electro--optic inclusions \cite{LM2006}
suggest the attractive possibility of fabricating porous SCMs with
sculptured--thin--film technology \cite{LakhtakiaB}.

\newpage

\end{document}